\documentclass{rrparticle}
\usepackage{graphicx}
\usepackage{amssymb}
\usepackage{dsfont}
\usepackage{color}

\title{Positive and negative mass solitons \\ in spin-orbit coupled Bose-Einstein condensates} 
\author[1]{V. Achilleos}
\author[1,*]{D.J. Frantzeskakis }
\author[2]{P.G. Kevrekidis }
\author[3,4]{P. Schmelcher }
\author[3]{ \\ J. Stockhofe}
\affil[1]{Department of Physics, University of Athens, Panepistimiopolis,
Zografos, Athens 157 84, Greece }
\affil[2]{Department of of Mathematics and Statistics, University of
Massachusetts, \\ Amherst, MA 01003-9305, USA}
\affil[3]{ Zentrum f\"ur Optische Quantentechnologien,
Universit\"at Hamburg, \\ Luruper Chaussee 149, 22761 Hamburg, Germany}
\affil[4]{The Hamburg Centre for Ultrafast Imaging,
Luruper Chaussee 149, 22761 Hamburg, Germany \\
$^*$Corresponding author E-mail: {\em dfrantz@phys.uoa.gr}}


\begin{document}

\maketitle
\begin{abstract}
We present a unified description of different types of matter-wave solitons 
that can emerge in quasi one-dimensional spin-orbit coupled (SOC) Bose-Einstein condensates (BECs).
This description relies on the reduction of the original two-component Gross-Pitaevskii 
SOC-BEC model to a single nonlinear Schr\"{o}dinger equation, via a multiscale expansion method.
This way, we find approximate bright and dark soliton solutions, for attractive and 
repulsive interatomic interactions respectively, for different 
regimes of the SOC interactions. 
Beyond this, our approach also reveals ``negative mass'' regimes, where 
corresponding ``negative mass'' bright or dark solitons can exist for repulsive or attractive interactions, respectively. 
Such a unique opportunity stems from the structure of the excitation spectrum of the SOC-BEC.
Numerical results are found to be in excellent agreement with our 
analytical predictions. 
\end{abstract}

\section{Introduction}

The recent experimental realization of spin-orbit coupling (SOC) of neutral 
atoms in Bose-Einstein condensates (BECs)~\cite{spiel1,spiel3,expcol} 
and fermionic gases~\cite{fermi1,fermi2} paved the way for a new era of studies in  
cold atom physics. Indeed, these recent developments suggested the opportunity that many 
fundamental phenomena of condensed-matter or even high-energy physics can be 
simulated in cold atom systems. Striking features such as {\it Zitterbewegung} have 
already been observed in experiments~\cite{zbengels,zbspiel}, while 
proposals for the study of quantum spin Hall effect~\cite{hallkett}, 
topological quantum phase transitions in Fermi gases~\cite{topfermi}, 
or even chiral topological superfluid phases~\cite{2dSO}, have also been reported. 
It is thus clear that the perspectives are very promising: indeed, 
currently available experimental techniques allow the creation of versatile 
effective gauge potentials in ultracold atoms, Abelian or even non-Abelian, 
relevant not only to condensed-matter physics 
but also to the understanding of interactions between elementary particles 
(for relevant reviews on this subject cf. Refs.~\cite{dalib,revspiel}).

Although many effects related to SOC-BECs can be explored and understood 
in the linear (non-interacting) limit, there has also been much interest in the 
study of purely nonlinear effects and, in particular, in localized nonlinear excitations 
that may be supported by SOC. This way, topological structures such as skyrmions~\cite{skyrm}, 
vortices \cite{xuhan}, and Dirac monopoles \cite{dirac}, as well as 
coherent structures in the form of
dark \cite{brand,VAd} or gap~\cite{konotop}, and bright \cite{VAb,VAbc} solitons (for repulsive and attractive 
interatomic interactions, respectively) have been studied in SOC-BECs. 
Solitonic structures have already been 
investigated extensively in single- and multi-component BECs (see, e.g., the 
reviews~\cite{review,review1}); nevertheless, the above mentioned recent studies have shown 
that the presence of SOC enriches significantly the possibilities regarding the structural form of 
solitons, as well as their stability and dynamical properties. As an example, we note the 
recent work \cite{ourZB}, where beating dark-dark solitons, that occupy both energy bands of the spectrum of a SOC-BEC and may perform {\it Zitterbewegung} oscillations, were predicted to occur.  Interestingly, the solitonic structures studied in Ref.~\cite{ourZB}, are embedded families of bright, twisted or higher excited solitons inside a dark soliton which, as well, are only 
possible due to the structure of the energy spectrum of SOC-BECs. 
Moreover, phenomena stemming from the interplay of SOC with additional
manipulation capabilities available in BECs such as optical lattices or
Raman coupling fields
are presently under intense investigation. Canonical examples thereof
can be seen in the manifestation of dynamical instabilities of excitations in
the vicinity of a band gap~\cite{engels_rec} of optical lattices,
or in the demonstration of a Dicke-type quantum phase transition,
by changing the Raman coupling strength between a spin-polarized and
a spin-balanced phase, 
in analogy to quantum optics~\cite{natcom}.

It is the purpose of this work to provide a unified description of different types of solitons 
that may occur in either attractive or repulsive quasi one-dimensional (1D) SOC-BECs. 
Our methodology relies on the study of soliton solutions of an effective scalar nonlinear 
Schr\"{o}dinger (NLS) equation: this can be derived from the system of the two 
coupled Gross-Pitaevskii equations describing the SOC-BEC via a multiscale expansion 
method. We first find either dark or bright soliton solutions, 
for repulsive or attractive interactions, respectively, of amplitudes and widths 
depending on the SOC 
parameters. These solitons may be formed in either of the two regions of the 
excitation energy spectrum, 
where the lowest energy band has either a single- or a double-well structure; so-called
``striped'' solitons, in the region where the double-well of the spectrum is symmetric, are 
shown to be possible, too. The above mentioned types of solitonic structures 
will be called hereafter ``positive mass solitons'', to distinguish them from 
the possibility of ``negative mass'' ones introduced in the following. 

Indeed, interestingly enough, our analytical approach reveals the existence of 
``negative mass'' parametric regions, meaning that the sign of the kinetic energy term 
(the dispersion coefficient in the effective NLS model) becomes inverted. In such 
negative mass regions, we show that bright or dark solitons can be formed in SOC-BEC 
with repulsive or attractive interactions, respectively. Pertinent solitonic structures 
will be called hereafter ``negative-mass solitons''.
Notice that such solitons cannot exist in usual single-component BECs described by a scalar 
NLS model: there, bright (dark) matter-wave solitons are formed only for BEC with 
attractive (repulsive) interactions, respectively \cite{we}. Instead, bright solitonic 
structures, for instance, can only exist as {\it gap solitons} in repulsive BECs loaded 
in an optical lattice potential, where the effective mass 
also changes sign (becomes negative) at the edge of the first Brillouin zone \cite{MKO}; for a theoretical proposal of the extension of such states
in SOC-BECs, see, e.g.,~\cite{konotop}.
Thus, our results suggest an alternative method of ``dispersion management'' of matter-waves based 
on the SOC effect
(as previous proposals relied on the use of optical lattices \cite{MKO2}). 

\section{The model and its analytical consideration}

We consider the case of a quasi-1D SOC-BEC, confined in a trap with
longitudinal and transverse frequencies, $\omega_x$ and $\omega_{\perp}$, 
such that $\omega_x \ll \omega_{\perp}$. In the framework of mean-field theory, 
this system is described by the energy functional \cite{spiel1,ho}:

\begin{align}
\mathcal{E}\!=\! \mathbf{\Psi}^{\dagger} \mathcal{H}_0 \mathbf{\Psi} +
\frac{1}{2} \left( g_{11}|\psi_1|^4+g_{22}|\psi_2|^4
+2g_{12}|\psi_1|^2|\psi_2|^2 \right),
\label{hamfull}
\end{align} where $\mathbf{\Psi}\equiv (\psi_1, \psi_2)^T$, and the condensate wavefunctions
$\psi_{1,2}$ are the two pseudo-spin components of the BEC. Furthermore, 
the single particle Hamiltonian $\mathcal{H}_0$ in Eq.~(\ref{hamfull}) reads:

\begin{align}
\mathcal{H}_0=\frac{1}{2m}(\hat{p}_x+k_L\hat{\sigma}_z)^2+V_{\rm tr}(x)+\Omega\hat{\sigma}_x+\delta\hat{\sigma}_z,
\label{h0}
\end{align} where $\hat{p}_x=-i\hbar\partial_x$ is the momentum operator in the longitudinal direction,
$m$ is the atomic mass, and $\hat{\sigma}_{x,z}$ are the Pauli matrices.
The SOC terms are characterized by the wavenumber $k_L$ of the Raman laser
which couples the two components, and the strength of the coupling $\Omega$. Moreover, 
$\delta$ is an energy shift due to the detuning from Raman resonance. 
The external trapping potential $V_{\rm tr}(x)$, is assumed to be of the usual parabolic form,  
$V_{\rm tr}=(1/2)m\omega_x^2x^2$. Finally, the effective 1D coupling constants $g_{ij}$ 
are given by $g_{ij}=2\hbar\omega_{\perp}\alpha_{ij}$, where $\alpha_{ij}$ 
are the s-wave scattering lengths; below, we will present results for both attractive ($\alpha_{ij}<0$) and repulsive ($\alpha_{ij}>0$) interatomic interactions. 

Using Eq.~(\ref{hamfull}), we can obtain the following dimensionless equations of motion:

\begin{eqnarray}
i\partial_t \psi_1 &=& \left(-\frac{1}{2}\partial^2_x-ik_L\partial_x + V_{\rm tr}
+ \sigma\left(|\psi_1 |^2+\beta |\psi_2|^2 \right)  +\delta \right)\psi_1 + \Omega\psi_2, 
\label{GP1a} \\
i\partial_t\psi_2&=&\left( -\frac{1}{2}\partial^2_x+ik_L\partial_x +V_{\rm tr}
+\sigma\left(\beta|\psi_1 |^2 + |\psi_2|^2\right) -\delta \right)\psi_2 + \Omega \psi_1,
\label{GP1b}
\end{eqnarray} where energy, length, time and densities are measured in units of 
$\hbar \omega_\perp$,  
$a_\perp$ (which is equal to $\sqrt{\hbar/ m \omega_\perp}$), $\omega_\perp^{-1}$, and 
$\alpha_{11}$, respectively, and we have also used the transformations 
$k_L\rightarrow a_\perp k_L$, $\Omega \rightarrow \Omega/( \hbar\omega_\perp)$ 
and $\delta \rightarrow \delta/( \hbar\omega_\perp)$. Notice that  
we have also assumed that the intraspecies scattering lengths are identical, 
$\alpha_{11}=\alpha_{22}$; this way, parameter $\sigma$, defined as 
$\sigma \equiv {\rm sgn}(\alpha_{11})={\rm sgn}(\alpha_{22})= \pm 1$, 
corresponds to repulsive and attractive intra-species interactions, 
respectively, while $\beta=\alpha_{12}/|\alpha_{11}|$. 
Finally, the trapping potential in Eqs.~(\ref{GP1a})-(\ref{GP1b}) is now given by 
$V_{\rm tr}(x) = (1/2)\omega_{\rm tr}^2 x^2$, where $\omega_{\rm tr} = \omega_x/\omega_\perp \ll 1$. 
To a first approximation, when sufficiently weak, the trapping potential can be neglected. Thus, 
below we 
will only consider the homogeneous case, i.e., $V_{\rm tr}=0$ (nevertheless, we have checked 
that the results do not change qualitatively upon including the parabolic trap -- see also  Refs.~\cite{VAb,VAd}). 

Below we will use the following parameter values: 
$k_L=8$ for the Raman wavenumber, $\Omega \in [0, 120]$ for the normalized 
Raman coupling strength, which are relevant to the experiments of Ref.~\cite{spiel1}.
Finally, for our analytical considerations, $\beta$ will be treated as an arbitrary (positive) parameter.
In fact, for the hyperfine states of $^{87}$Rb commonly used in today's SOC-BEC experiments this parameter takes values 
$\beta \approx 1$, due to the fact that the inter- and intra-species scattering lengths take approximately the same values.
For this reason, in the simulations that we will present below, this parameter is fixed to the value $\beta=1$; 
nevertheless, we have confirmed that physically relevant deviations from this value do not alter 
qualitatively the results.

We now use a multiscale perturbation method to reduce the system of Eqs.~(\ref{GP1a})-(\ref{GP1b}) 
to a scalar NLS equation. Such a reduction will allow us to derive approximate analytical 
bright and dark soliton solutions, depending on the type of the interactions.
To proceed, we introduce the ansatz:

\begin{align}
\mathbf{\Psi}=\sum_{n=1}^{\infty}\epsilon^n\mathbf{u}_ne^{i(kx-\mu t)}\equiv
\sum_{n=1}^{\infty}\epsilon^n 
\left(\begin{array}{c}
U_n \\ V_n
\end{array}
\right)\phi_ne^{i(kx-\mu t)}, 
\label{aexpansion}
\end{align} where the vectors $\mathbf{u}_n=[U_n,V_n]^T\phi_n$ are composed by the coefficients $U_n$ and $V_n$
and the unknown field envelopes $\phi_n\equiv \phi_n(T,X)$. The latter are assumed to 
be functions of the slow variables $T=\epsilon^2t$ and $X=\epsilon(x-vt)$, where $v$ is the 
velocity (to be determined). Additionally,
$k$ is the momentum, $\mu=\omega+\epsilon^2\omega_0$ the chemical potential; here, 
$\omega$ is the energy in the linear limit, $\epsilon^2\omega_0$ 
is a small deviation about this energy ($\epsilon \ll 1$), and $\omega_0/\omega=\mathcal{O}(1)$.  
Introducing Eq.~(\ref{aexpansion}) into Eqs.~(\ref{GP1a})-(\ref{GP1b}), we derive the following 
set of equations at orders $\mathcal{O}(\epsilon)$, $\mathcal{O}(\epsilon^2)$ and 
$\mathcal{O}(\epsilon^3)$, respectively:

\begin{align}
\mathbf{W}\mathbf{u}_1&=0, \label{ord1} 
\\
\mathbf{W}\mathbf{u}_2&=i\mathbf{W_0} \partial_X\mathbf{u}_{1}, \label{ord2} 
\\
\mathbf{W}\mathbf{u}_3&=i\mathbf{W_0}\partial_X\mathbf{u}_{2} 
- \left(i\partial_{T}+\frac{1}{2}\partial_{X}^2 -\mathbf{A}+\omega_0\right)\mathbf{u}_{1},
\label{ord3}
\end{align} where $\mathbf{W_0} = (\mathbf{W}- \omega \mathds{1})'$ (primes denote hereafter differentiation 
with respect to $k$), while matrices $\mathbf{W}$ 
and $\mathbf{A}$ are:

\begin{align}
\mathbf{W}&=\left [ \begin{array}{c c}
\omega-k^2/2-kk_L-\delta & -\Omega \\
-\Omega & \omega-k^2/2+kk_L+\delta
\end{array}
\right], 
\\
\mathbf{A}&=\left [ \begin{array}{c c}
\sigma\left(|U_1|^2 +\beta |V_1|^2\right) & 0 \\
0 & \sigma\left(\beta|U_1|^2 + |V_1|^2\right)
\end{array}
\right]. 
\end{align}

\begin{figure}[tbp]
\centering
\includegraphics[width=8.7cm]{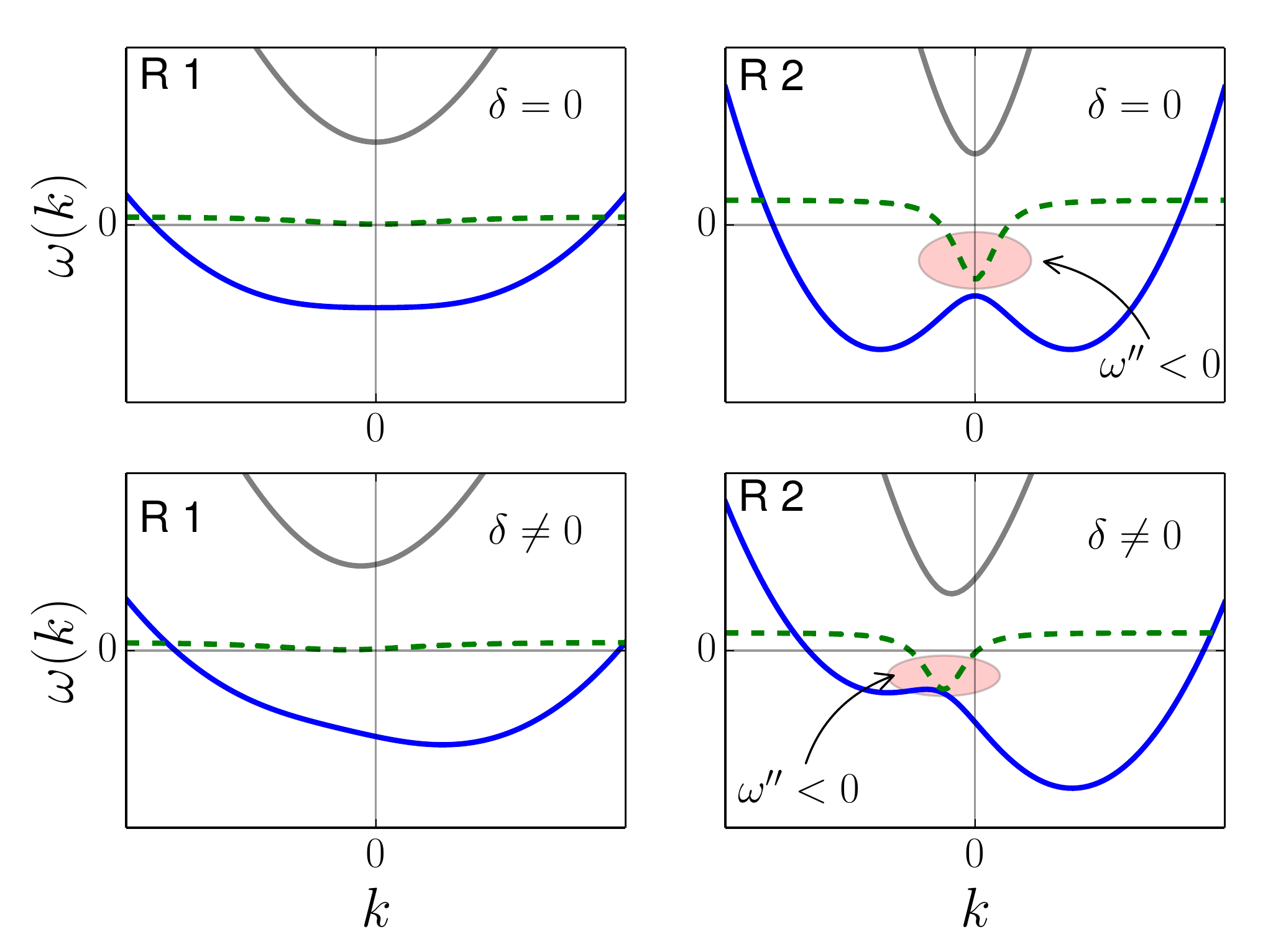}
\caption{(Color online) Sketches of the linear energy spectrum $\omega(k)$ 
for Region~1 and Region~2, denoted by R~1 and R~2; cf. left and right panels, respectively. 
Top (bottom) panels correspond to $\delta=0$ ($\delta \ne 0$), and dashed 
lines depict the group velocity dispersion $\omega^{\prime\prime}(k)$ of the lower band.}
\label{fig1}
\end{figure}

At $\mathcal{O}(\epsilon)$, the solvability condition 
${\rm det}\mathbf{W}=0$ yields the linear excitation energy spectrum

\begin{equation}
\omega = \omega_{\pm}(k) = \frac{1}{2} k^2 \pm\sqrt{(kk_L+\delta)^2+\Omega^2}, 
\label{dr}
\end{equation} which consists of two different bands, the upper and lower one, as shown in Fig.~\ref{fig1}.

The lower band of the spectrum $\omega_-$ has a single local minimum for $\Omega/k_L^2>1$, 
and two local minima for $\Omega/k_L^2<1$; respective regions of the parameter space will be 
called hereafter ``Region~1'' and ``Region~2'' respectively (cf. left and right panels of 
Fig.~\ref{fig1}). Notice that for $\delta=0$ the spectrum is symmetric with respect
to $k=0$, while for $\delta\ne 0$ this symmetry is broken (cf. top and bottom 
panels of Fig.~\ref{fig1}). We should also mention, at this point, that below we 
will focus on nonlinear states, in the form of matter-wave solitons, corresponding 
to the lower energy band (ground state). However, as shown in Ref.~\cite{ourZB}, 
solitons may also exist in an excited state occupying simultaneously both energy bands \cite{note}. 
Finally, at the same order $\mathcal{O}(\epsilon)$, 
the compatibility condition of Eq.~(\ref{ord1}) yields:

\begin{equation}
\mathbf{u}_1=\left(\begin{array}{c} 1 \\  Q
\end{array}    \right)\phi_1(X,T), 
\label{u1}
\end{equation}
where the parameter $Q$ is given by:   

\begin{equation}
Q=Q(\omega,k) \equiv \frac{1}{\Omega}\left(\omega-\frac{1}{2}k^2 -kk_L-\delta\right). 
\label{Q}
\end{equation}

Notice that the above parameter sets the left and right eigenvectors of $\mathbf{W}$ 
at eigenvalue $0$, given by $\mathbf{L}=[1,Q]$ and $\mathbf{R}=[1,Q]^T$, respectively. 

Next, at the order $\mathcal{O}(\epsilon^2)$, the compatibility condition of Eq.~(\ref{ord2}), 
$\mathbf{L}\mathbf{W_0}\mathbf{R}=0$, 
fixes $v$ at the group velocity, $v=\omega'(k)$, yielding the explicit condition 

\begin{equation}
v=k -k_L\frac{Q^2-1}{Q^2+1}.
\label{vgr}
\end{equation}

This means that solutions for $\phi$ in the lower energy band 
may be either {\it static} for $k=k_{\rm min}$ 
(choosing a minimum $k_{\rm min}$ of the dispersion where the group velocity vanishes), 
or {\it moving} for $k \ne k_{\rm min}$. 
Note that, in the following, both static and moving soliton solutions will be presented. 
At the same order, we also obtain the form of the solution for $\mathbf{u}_2$:

\begin{equation}
\mathbf{u}_2=-i\left(\begin{array}{c} 1 \\  Q'
\end{array}    \right)\partial_X\phi_1(X,T).
\label{u2}
\end{equation}

Finally, the compatibility condition for Eq.~(\ref{ord3}) at 
$\mathcal{O}(\epsilon^3)$, together with Eqs.~(\ref{u1})
and (\ref{u2}), yields the following scalar NLS equation for the unknown field $\phi_1$:

\begin{align}
i\partial_T\phi=-\frac{1}{2}\omega''\partial_{X}^2\phi+\sigma\nu|\phi|^2\phi-\omega_0\phi.
\label{nls1}
\end{align} where the subscript has been omitted for convenience, while parameter $\nu$ is given by:

\begin{align}
\nu(k)=\frac{Q^4+2\beta Q^2+1}{1+Q^2}, 
\label{nu}
\end{align} and it is always positive, since parameter $Q$ [cf. Eq.~(\ref{Q})] is real 
(note that here we have assumed that $\beta>0$). 
The above NLS model plays a central role in our analysis, as it will be used 
below for the derivation of various soliton families.

\section{Soliton solutions}

The sign of the nonlinearity and dispersion coefficients of the above effective NLS model
is crucial in determining the types of solitons that can be supported: 
for $\omega'' \sigma >0$ the solitons are dark, while for $\omega'' \sigma <0$ the 
solitons are bright. Obviously, the sign of the nonlinearity 
coefficient depends only on the type of interatomic interactions: 
$\sigma =\pm 1$ for repulsive and attractive interactions, respectively. 

On the other hand, let us take a closer look at the dispersion coefficient 
$\omega'' \equiv \partial^2 \omega/\partial k^2$, which can also be viewed as the inverse 
of the effective mass, i.e., $\omega'' \propto m_{\rm eff}^{-1}$ (see, e.g., Refs.~\cite{MKO,MKO2}). 
The dependence of this coefficient on momentum $k$ is depicted by the dashed lines 
in Fig.~\ref{fig1}. It is observed that in the lower energy band, in both Regions~1 and~2,  
and for $k=k_{\rm min}$ (where group velocity is zero, as mentioned above), one has always  
$\omega''>0$. This indicates the existence of ``positive mass'' stationary dark (bright) solitons  
in SOC-BECs with repulsive (attractive) interactions. Positive mass moving solitons can exist 
as well, featuring a finite group velocity that can be found for $k \ne k_{\rm min}$. 
Notice that dark solitons exist inside the linear band ($\omega_0>0$), 
while the bright solitons are found in the infinite gap below the lower 
energy band ($\omega_0<0$); this can be readily seen by the stationary 
form of Eq.~(\ref{nls1}), corresponding to $\partial_T\phi=0$

On the other hand, in Region~2, there exist intervals 
of momenta (depicted by the ellipses in the right panels 
of Fig.~\ref{fig1}) where $\omega''<0$. Thus, also ``negative mass'' moving dark (bright) solitons  
can exist in SOC-BECs with attractive (repulsive) interactions. 

We now proceed by presenting different types of exact soliton solutions of Eq.~(\ref{nls1}) 
and corresponding approximate soliton solutions of the original system of Eqs.~(\ref{GP1a})-(\ref{GP1b}). 
We will also study the solitons' properties 
by means of direct numerical simulations.  

\subsection{Positive mass solitons}

We start with the case of positive mass solitons, which may be either stationary (exhibiting 
a momentum $k=k_{\min}$ at the minima of the energy spectrum where 
the group velocity $v$ vanishes) or moving (for momenta $k\ne k_{\min}$), as mentioned above. 

For repulsive interactions, $\sigma>0$, we may write the dark soliton solution of Eq.~(\ref{nls1}), with $\omega_0>0$, in the form \cite{review,we}:

\begin{align}
&\phi_{\rm d}=\sqrt{\omega_0/\nu}\left(\cos\theta{\rm tanh}z_d+i \sin\theta\right), 
\label{sold1} 
\end{align} where $z_d=\sqrt{\omega_0/\omega''}\cos\theta[X-X_0(T)]$;
here, $X_0(T)$ is the soliton center, while the so-called soliton phase angle $\theta$ 
controls the soliton amplitude, $\sqrt{\omega_0/\nu}\cos\theta$, and soliton velocity 
in the slow time scale $T$ through the equation $dX_0/dT=\sqrt{\omega_0/\omega''}\sin\theta$.  
Notice that the above soliton solution is characterized by two free parameters, one 
for the background, $\omega_0$, and one for the soliton, $\theta$.

For attractive interactions, $\sigma>0$, a bright soliton solution of 
Eq.~(\ref{nls1}), with $\omega_0<0$, may be written as:

\begin{align}
&\phi_{\rm b}=\eta{\rm sech}z_b \exp(i\kappa X),
\label{solb1}
\end{align} where $z_b=\eta\sqrt{\nu/\omega''}[X-X_0(T)]$; here, $\eta$ is the soliton amplitude which 
is connected with the parameter $\omega_0$ through the relation 
$\omega_0=(\kappa^2+\eta^2\nu/\omega'')/2$, $X_0(T)$ is the soliton center, 
while the wavenumber $\kappa$ is connected with the soliton velocity through  $dX_0/dT=\omega''\kappa$. 
As in the previous case, the above soliton solution has two free parameters, 
$\omega_0$ and $\kappa$.

Having found the solutions 
of Eq.~(\ref{nls1}), we may write the approximate stationary soliton solutions 
for the two components
of the SOC-BEC, in the original coordinates as follows:

\begin{align}
&\left(
\begin{array}{c}
\psi_1 \\
\psi_2
\end{array}
\right) \approx \left(
\begin{array}{c}
1 \\
Q
\end{array}
\right)\epsilon\sqrt{\omega_0/\nu}\left(\cos\theta{\rm tanh}z_d+i \sin\theta\right) 
\exp[ikx -i(\omega+\epsilon^2\omega_0)t], 
\label{sol1a} \\
&\left(
\begin{array}{c}
\psi_1 \\
\psi_2
\end{array}
\right) \approx \left(
\begin{array}{c}
1 \\
Q
\end{array}
\right)\epsilon\eta{\rm sech}z_b \exp[i(k+\epsilon \kappa)x -i(\omega-\epsilon^2\omega_0)t], 
\label{sol1b}
\end{align} for the dark and bright solitons respectively; recall that $Q$ depends on $k$ [or $\omega$, 
as per the dispersion relation~(\ref{dr})].

Another family of soliton solutions can also be found as follows.
In Region~2, and for $\delta=0$ (see top right panel of Fig.~\ref{fig1}), there exist two 
degenerate minima at specific values of momenta, say $k_c$ and $-k_c$, for which 
two different nonlinear solutions can be obtained, as per the asymptotic analysis presented above. 
However, in the linear limit, a superposition of these states is also a solution 
with the same energy. Using a continuation argument, it turns out to be 
possible to find a nonlinear solution as well. In the case $\delta=0$ and $\beta=1$, we may employ the symmetry of Eqs.~(\ref{GP1a})-(\ref{GP1b}), namely $\psi_1=-\bar{\psi}_2$ (bar denotes complex conjugation), 
and find approximate solutions of the original equations in the form:

\begin{align}
\left( 
\begin{array}{c}
\psi_1 \\
\psi_2
\end{array}
\right) \approx \epsilon C \phi_{d,b}
\left( 	
\begin{array}{c}
q_{+} \cos(k_{\rm c} x) 
+i q_{-} \sin(k_{\rm c} x) \\
-q_{+} \cos(k_{\rm c} x) 
+i q_{-}\sin(k_{\rm c} x)
\end{array}
\right), 
\label{sol3}
\end{align} where $q_{\pm} = \Omega^{-1}+Q(\pm k_{\rm c})$ and $C$ is a free parameter. 

\begin{figure}[tbp]
\centering
\includegraphics[width=12cm]{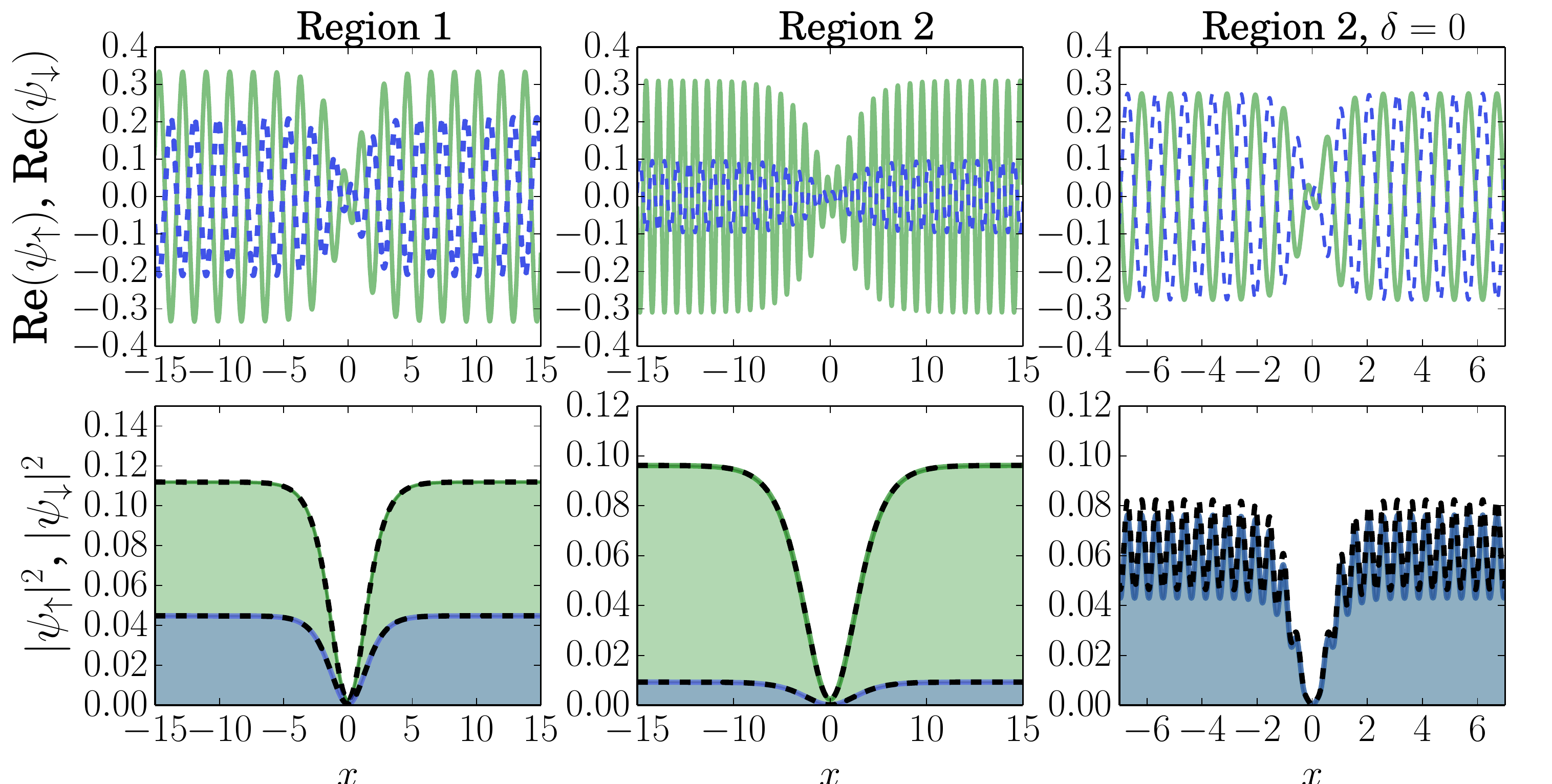}
\caption{(Color online) Positive mass dark solitons for $\sigma = +1$. 
In the top panels shown are the real parts of $\psi_{1,2}$,
while the bottom panels depict the densities of the two components. The left
column corresponds to Region~1 with $\Omega/k_L^2=1.56$ and the middle
panel to Region~2 with $\Omega/k_L^2=0.78$ both for $\delta=20$. The right panels
depict the ``stripe'' solitons for $\Omega/k_L^2=0.78$ and $\delta=0$.
Dashed (black) lines in the bottom panels depict corresponding analytical results.}
\label{fig2}
\end{figure}

\begin{figure}[tbp]
\centering
\includegraphics[width=12cm]{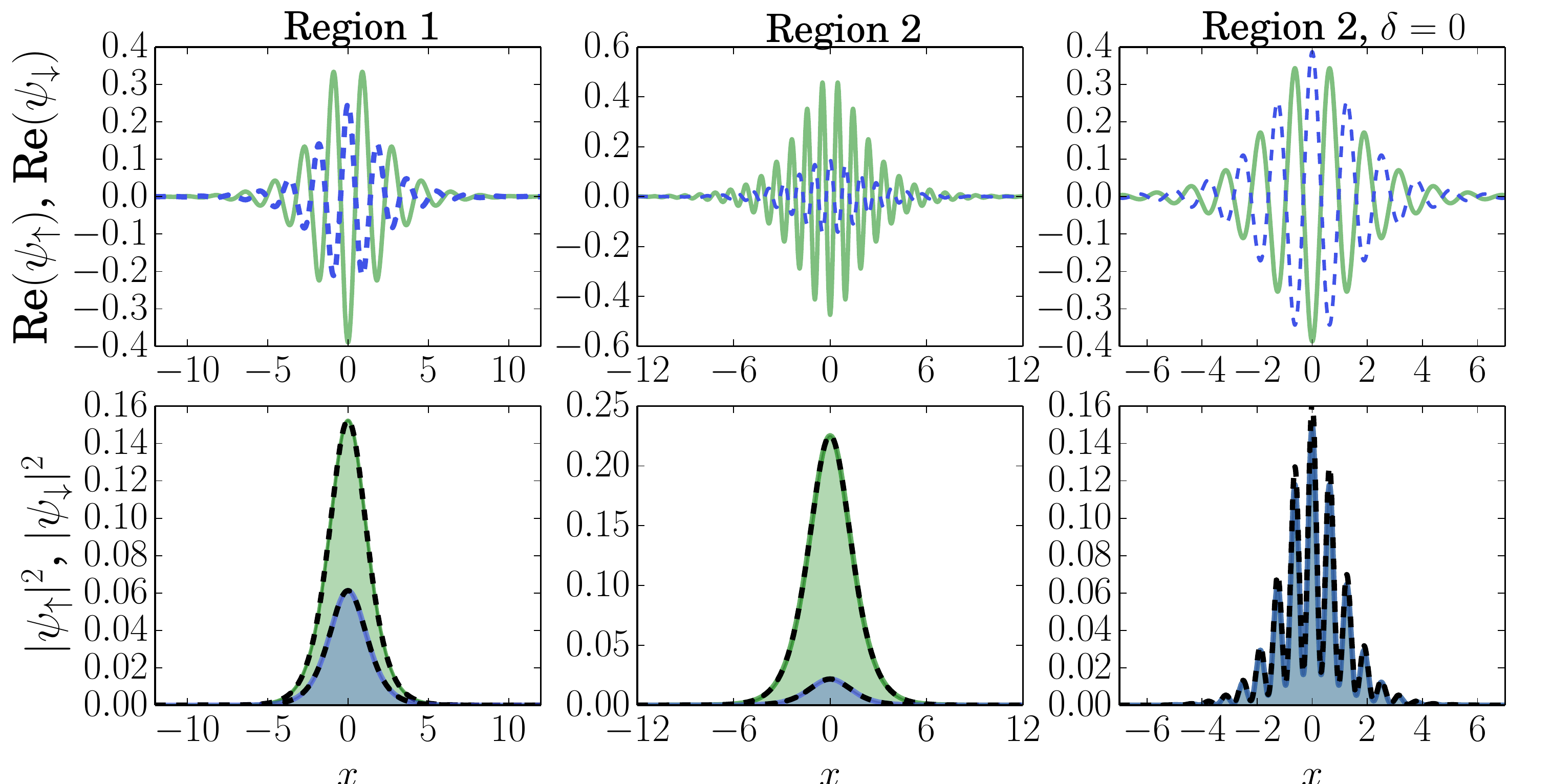}
\caption{(Color online) Positive mass bright solitons for $\sigma = -1$. 
In the top panels shown are the real parts of $\psi_{1,2}$,
while the bottom panels depict the densities of the two components. The left
column corresponds to Region 1 with $\Omega/k_L^2=1.56$ and the middle
panel to Region~2 with $\Omega/k_L^2=0.78$ both for $\delta=20$. The right panels
depict the ``stripe'' solitons for $\Omega/k_L^2=0.78$ and $\delta=0$. Dashed (black)
lines in the bottom panels depict corresponding analytical results.}
\label{fig3}
\end{figure}

Examples of all three different families of dark and bright soliton solutions 
are shown in Figs.~\ref{fig2} and \ref{fig3}, respectively. These states were 
obtained numerically, by solving (via a fixed point algorithm) 
the stationary form of Eqs.~(\ref{GP1a})-(\ref{GP1b}) obtained by using 
$\mathbf{\Psi}(x,t)=\mathbf{\Phi}(x)\exp(-i\mu t)$; then, for $\mathbf{\Phi}(x)$ 
the stationary form of the solitons of Eqs.~(\ref{sol1a}), (\ref{sol1b}) and (\ref{sol3}) 
was used as an initial guess. Notice that for obtaining these solitons, we have used 
the momentum $k=k_{\rm min}$, corresponding to zero group velocity. In other words, the 
solitons shown in Figs.~\ref{fig2} and \ref{fig3} are stationary ones; for arbitrary momenta,  positive mass {\it moving} solitons also exist, but we will not discuss them further 
(they were found to exist and be stable in parameter regions where their stationary 
counterparts are stable \cite{VAd,VAb}).

In the left panel of Fig.~\ref{fig2} (Fig.~\ref{fig3}), a dark (bright) soliton 
in Region~1 is shown for $\Omega/k_L^2=1.56$ and $\delta=20$. In the middle panel 
of these figures, we show respective solitons in Region~2, for $\Omega/k_L^2=0.78$ 
and $\delta=20$. Note that the number of atoms 
of the two components is generally different, with the asymmetry arising due to the 
presence of the constant factor $Q$ [cf. Eqs.(\ref{sol1a})-(\ref{sol1b})]. 
Finally, in the right panel of Fig.~\ref{fig2} (Fig.~\ref{fig3}), an example of 
a ``stripe'' dark (bright) soliton is shown for $\Omega/k_L^2=0.78$ and $\delta=0$. 
Note that in this case the two components have the same amplitudes
and number of atoms, which is a consequence of the symmetry of these states, 
associated with the fact that $\psi_1=-\bar{\psi}_2$.
In both figures~(Figs.~\ref{fig1} and~\ref{fig2}), we also plot the analytical result 
of Eqs.~(\ref{sol1a})-(\ref{sol3}), depicted by a dashed (black) line, illustrating the very 
good agreement between our approximate analytical result and the numerically obtained solutions. 

\subsection{Negative mass solitons}

We now proceed with the case of negative mass solitons, which can be found in specific 
domains of Region~2, where $\omega''<0$ (cf. ellipses in the right panels of Fig.~\ref{fig1}). 
In such a case, it is possible to obtain bright soliton solutions for a {\it repulsive} SOC-BEC, 
and dark soliton solutions for an {\it attractive} BEC. This is a special feature of the spin-orbit 
coupling and of the resulting structure of the linear energy spectrum.
In fact, this is a feature typically {\it absent} from homogeneous
(especially single component) systems, although such a possibility
can be generated in inhomogeneous BECs, e.g. in the presence
of an optical lattice~\cite{MKO}. 
Negative mass bright and dark soliton solutions have the same functional form as positive mass ones  
[Eqs.~(\ref{sol1a})-(\ref{sol1b})], but now for $\sigma=+1$ and $\sigma=-1$, respectively.

\begin{figure}[tbp]
\centering
\includegraphics[width=13cm]{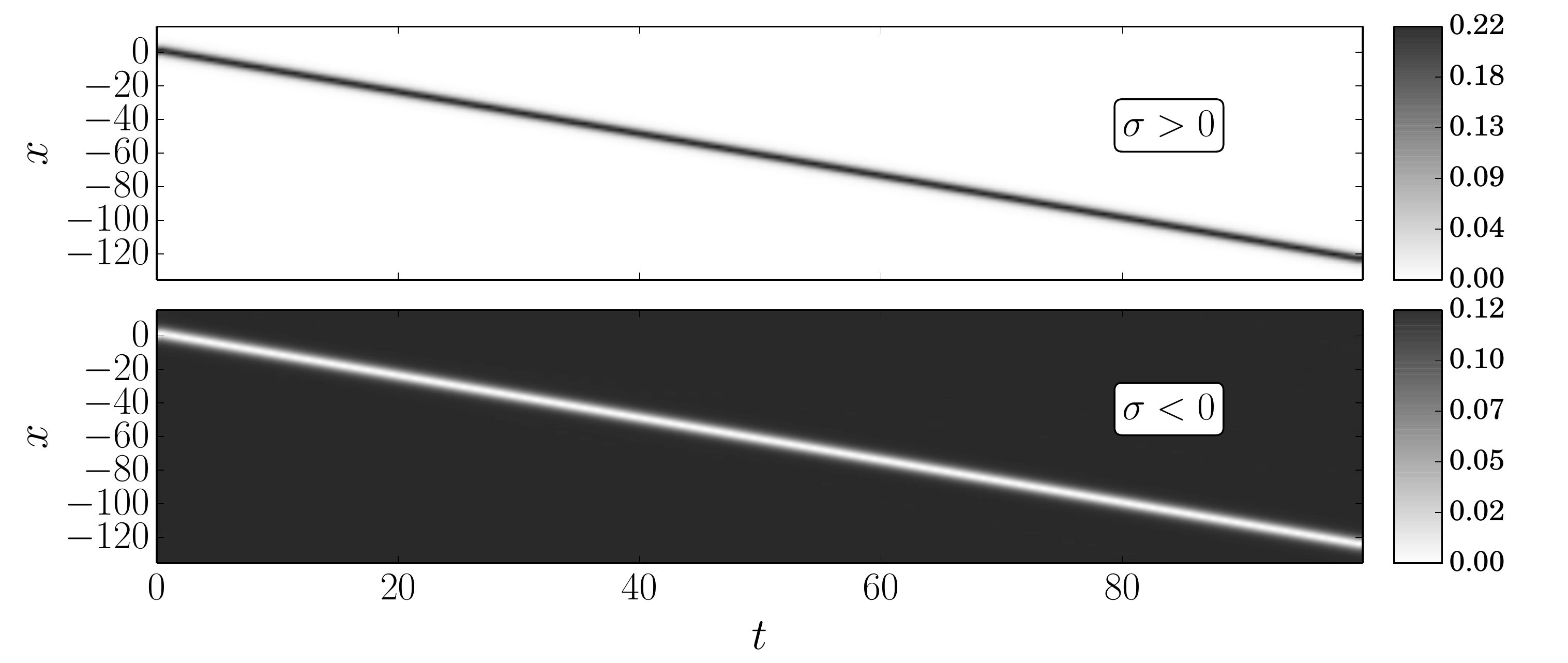}
\caption{Top panel: contour plot showing the evolution of the total density
$|\psi_1|^2+|\psi_2|^2$, for an initial bright soliton in the case of repulsive interactions ($\sigma=+1$).
Bottom panel: same as above, but now for a dark soliton in the case of attractive interactions ($\sigma=-1$). Parameter values used in both cases are 
$\Omega/k_L^2=0.49$, $\delta=4$, $k=-2.3$ and $\epsilon^2\omega_0=0.1$.}
\label{fig4}
\end{figure}

In order to illustrate the existence and dynamics of such states, we numerically integrate 
Eqs.~(\ref{GP1a})-(\ref{GP1b}) with an initial condition corresponding to a bright soliton in the 
repulsive case and to a dark soliton for the attractive case. 
The evolution of a bright soliton propagating in a SOC-BEC with repulsive interactions, 
is shown in the top panel of Fig.~\ref{fig4}, with parameter values $\Omega/k_L^2=0.49$, 
$\delta=4$, $\epsilon^2\omega_0=0.1$ and a momentum $k=-2.3$.
On the other hand, the evolution of a dark soliton, but for an attractive SOC-BEC, 
is shown in the bottom panel of the same figure, with the same parameter values as above. 
In both cases the solitons are found to be robust, and propagate without any distortion 
or significant emission of radiation; in fact, we have confirmed their robustness for times 
up to at least $t=400$ in dimensionless units.
Notice that in the numerical simulations we have also checked that the plane wave background of 
a negative mass dark soliton ($\sigma=-1$) is not subject to the modulational instability: in 
the simulations, we used a perturbation of wavenumber $K$ outside the negative-mass region (e.g., 
with $K=k_{\rm min}$) and found that the soliton background remains modulationally stable. 

\begin{figure}[tbp]
\centering
\includegraphics[width=13cm]{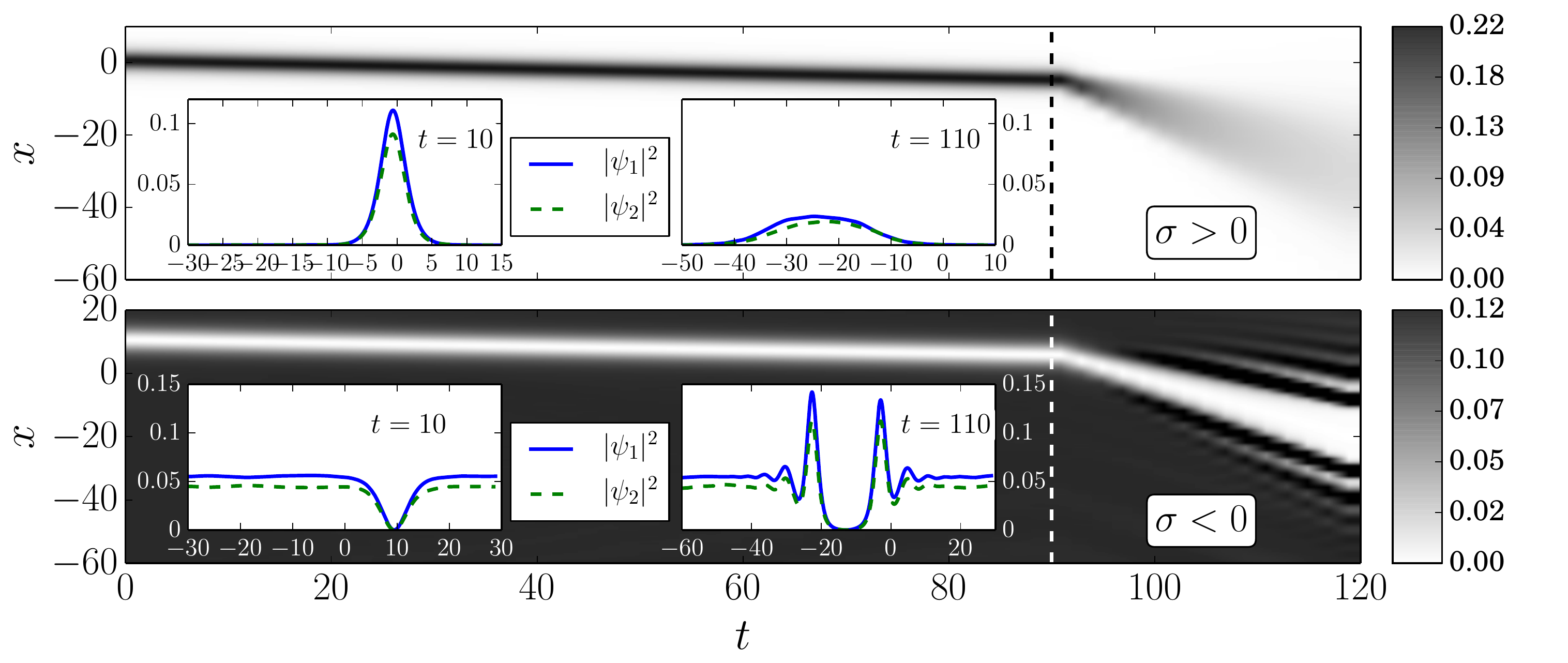}
\caption{(Color online) Same as in Fig.~\ref{fig4} but for $k=-0.9$.
The vertical dashed lines indicates the instant when the spin-orbit coupling is switched off and the 
insets show each component's density at $t=10$ (left) and $t=110$ (right).}
\label{fig5}
\end{figure}

The negative mass solitons shown in Fig.~\ref{fig4} feature a relatively large velocity: 
for a typical parameter value of the transverse frequency 
$\omega_\perp\approx 2\pi\times 400$~Hz for Rubidium atoms, 
the solitons propagate about $65$ microns in $0.4$s (i.e., the velocity is 
$|v|=162~\mu$m/s or $|v|=1.2$ in dimensionless units). Nevertheless, slower 
(and even stationary) solitons can also be obtained upon properly 
choosing different momenta values. A pertinent example is shown in the top panel of 
Fig.~\ref{fig5}: there, a bright soliton of momentum $k=-0.9$, corresponding
to a group velocity $v=-0.05$, is depicted. 

For the above case of slower solitons, and to better illustrate the fact 
that negative mass solitons exist only due to the presence of spin-orbit coupling, 
we perform the following numerical experiment (cf. Fig.~\ref{fig5}).
At time $t=90$, we turn off the SOC terms upon setting $\Omega,k_L$ and $\delta$ equal to zero. 
Then, as it is expected for a condensate with repulsive interactions, the bright soliton 
is not supported and, as a result, it begins to spread. The insets show the density profile 
of the two components $|\psi_{1,2}|^2$ when spin-orbit coupling is present (left) and when it
is turned off (right).

We have also studied the case of a negative mass dark soliton, 
propagating in an attractive BEC. A pertinent example is shown in the bottom panel of 
Fig.~\ref{fig5}, for the same parameter values (the dark soliton also has the 
same group velocity as its bright counterpart). 
After the initial undisturbed propagation of the soliton, we again switch off the spin-orbit coupling, 
at $t=90$. Then, as it is natural for the attractive case, 
the dark soliton can no longer be supported and is destroyed (its associated density
dip spreads), while the background on which it was supported becomes
subject to a modulational instability, gradually producing an array
of bright solitary waves.

Note that without switching off the SOC terms, we have confirmed the robust evolution 
of these ``slow'' negative mass bright and dark solitons for the repulsive and attractive case, 
respectively, for up to $t=400$ in dimensionless units.

\section{Conclusions}

Concluding, in this work we have provided a unified description of different types of solitons 
that may occur in quasi-1D SOC-BECs for both repulsive and attractive interactions. Our 
approach relies on a multiscale expansion method, which was used to reduce the original
system of two coupled Gross-Pitaevskii equations into an effective scalar nonlinear Schr\"{o}dinger (NLS) equation. 

Investigating the dispersion coefficient of the effective NLS model, we identified regions 
where it can become either positive or negative. This important effect, which arises 
only due to the presence of spin-orbit coupling, is reminiscent of what happens in the case of 
single-component BECs trapped in optical lattices: in the latter setting, due to the possibility of change of sign of the 
dispersion (effective mass in the framework of the relevant Gross-Pitaevskii model), bright 
(gap) solitons can be formed in BEC with repulsive interactions \cite{MKO}. This suggests the use of SOC for 
dispersion management of matter-waves, as in the case of BECs loaded in optical lattices \cite{MKO2}. 

Employing the above finding, we presented both positive and negative mass solitons that 
can be formed for positive and negative dispersion coefficient, respectively. 
In particular, positive mass dark (bright) solitons were found for repulsive 
(attractive) interactions, in either of the two regions of the energy 
spectrum, where the lowest energy band has either a single- or a double-well structure. 
Additionally, it was shown that ``striped'' solitons, in the region where the double-well of the 
spectrum is symmetric, are possible, too. These positive mass solitons may 
exist in either stationary or moving form. 

On the other hand, we have shown that negative mass bright (dark) solitons can be formed 
in SOC-BEC with repulsive (attractive) interactions; these solitons can also exist 
either as stationary or moving objects. Using direct numerical simulations, 
we found that such negative mass solitons are quite robust and can only exist in the presence 
of spin-orbit coupling. Indeed, letting the solitons evolve up to a certain time 
and then switching off the SOC terms, we found that they are not supported by the system without these terms.

Our methodology and findings suggest many interesting future research directions. First, it would 
be interesting to further investigate and propose mechanisms of SOC-mediated dispersion 
management of matter-waves. Additionally, it would be relevant to generalize our results for 
multi-component, $F=1$ spinor SOC-BEC \cite{Lan2014} and study solitons in such systems. 
Another interesting theme deserving attention is the analysis of modulational instability of plane waves 
in the original system of Eqs.~(\ref{GP1a})-(\ref{GP1b}); such an analysis, apart from being interesting by 
itself, would be particularly relevant to the stability of dark solitons in this setting. 
Finally, it would be particularly 
interesting to extend the above considerations to higher-dimensional settings
and explore the dynamics, e.g., of vortices or vortex rings in the presence of
such effective dispersions.

\emph{Acknowledgements.} The work of V.A. and D.J.F. was partially supported by the Special Account for 
Research Grants of the University of Athens. V.A., D.J.F. and P.G.K. appreciate warm 
hospitality at the University of Hamburg, where part of this work was carried out. P.G.K. acknowledges support from the National Science Foundation
under grants CMMI-1000337, DMS-1312856, from FP7-People under grant
IRSES-605096, from the Binational
(US-Israel) Science Foundation through grant 2010239 and
from the US-AFOSR under grant FA9550-12-10332. 
J.S. acknowledges support from the Studienstiftung des deutschen Volkes.


\begin{thebibliography}{99}
\bibitem{note} The excitation of the upper band was demonstrated in the experiments of 
Refs.~\cite{zbengels,zbspiel}, where Zitterbewegung oscillations were observed. Similarly, 
solitons occupying both upper and lower bands may, under certain conditions, exhibit 
Zitterbewegung oscillations as well \cite{ourZB}.
\end{thebibliography}
\end{document}